\documentclass[12pt,preprint]{aastex}
\def\psr{PSR~J1740$-$5340}
\def\com{COM~J1740$-$5340}

\def\ltsima{$\; \buildrel < \over \sim \;$}
\def\gtsima{$\; \buildrel > \over \sim \;$}
\def\lsim{\lower.5ex\hbox{\ltsima}}
\def\gsim{\lower.5ex\hbox{\gtsima}}
\def\lapp{\ifmmode\stackrel{<}{_{\sim}}\else$\stackrel{<}{_{\sim}}$\fi}
\def\gapp{\ifmmode\stackrel{>}{_{\sim}}\else$\stackrel{<}{_{\sim}}$\fi}

\newdimen\minuswidth    %define @ width of minus sign for tables
\setbox0=\hbox{$-$}
\minuswidth=\wd0
\catcode`@=\active
\def@{\kern\minuswidth}
\newdimen\digitwidth    %define ! a one digit width for tables
\setbox0=\hbox{\rm0}
\digitwidth=\wd0
\catcode`!=\active
\def!{\kern\digitwidth}
 
\begin{document} 
 
\title{The complex  H-$\alpha$ line profile of
the bright companion to PSR J1740-5340 in NGC 6397.
\footnote{Based on observations collected at the
European Southern Observatory, Chile, proposal 69.D-0264}
}

\author{
E. Sabbi\footnote{Dipartimento di Astronomia Universit\`a 
di Bologna, via Ranzani 1, I--40127 Bologna, Italy,
sabbi@bo.astro.it, ferraro@bo.astro.it},
R. Gratton
\footnote{INAF--Osservatorio Astronomico di Padova,
vicolo dell'Osservatorio 5, I--35122 Padova, Italy,
gratton@pd.astro.it},
F.R. Ferraro\altaffilmark{2}, 
A. Bragaglia\footnote{INAF--Osservatorio Astronomico di Bologna, via
Ranzani 1, I--40127 Bologna, Italy, angela@bo.astro.it},
A. Possenti\footnote{INAF--Osservatorio Astronomico di Cagliari,
Loc. Poggio dei Pini, Strada 54, I--09012 Capoterra, Italy,
possenti@ca.astro.it},
N. D'Amico\footnote{INAF--Osservatorio Astronomico di Cagliari,
Loc. Poggio dei Pini, Strada 54, I--09012 Capoterra, Italy, 
\& Dipartimento di Fisica, Universit\`a di Cagliari,
Strada Provinciale Monserratu--Sestu, km 0.700, I--09042 Monserrato, Italy,  
damico@ca.astro.it},
F. Camilo\footnote{Columbia Astrophysics Laboratory, Columbia
University, 550 West 120th Street, New York, NY 10027, 
fernando@astro.columbia.edu}
}
\medskip

\begin{abstract}
We present a detailed study of the H$\alpha$ and He~{\sc i} spectral
features of \com~(the companion to \psr~in the Galactic Globular
Cluster NGC~6397), exploiting a series of high resolution spectra
obtained at different orbital phases. The H$\alpha$ absorption line
shows a complex two-component structure, revealing that optically thin
hydrogen gas resides outside the Roche lobe of \com. The line
morphology precludes the existence of any residual disk around the
millisecond pulsar, and suggests the presence of a stream of material
going from the companion toward the neutron star. This material never
reaches the neutron star surface, being driven back by the pulsar
radiation far beyond \com. By analyzing the He~{\sc i} absorption
lines as a function of orbital phase, we infer the presence of an
overheated longitudinal strip (about 150 times narrower than it is
long) on the \com~surface facing the radio pulsar.
\end{abstract}


\section{Introduction} 
\label{intro}
 
The nearby Galactic globular cluster NGC~6397 hosts an eclipsing binary
millisecond pulsar (\psr; D'Amico et al. 2001a), whose companion (\com) has
been optically identified (Ferraro et al. 2001) with a bright (V$\sim$16.6)
variable star which has completely filled its Roche lobe (Ferraro et al.
2003).  Radio (D'Amico et al. 2001b) and X-ray observations (Grindlay et al.
2001, 2002) show evidence of interaction between the millisecond pulsar (MSP)
and mass released from the companion. Interestingly, despite the large pulsar
rotational energy loss, no photometric signature of heating can be found in the
light curves (Kaluzny et al. 2003, Orosz \& van~Kerkwijk 2003), which are
instead dominated by ellipsoidal variations (Orosz \& van~Kerkwijk 2003). The
combination of these observational facts represents a fascinating mosaic that
has triggered a wealth of hypotheses about the origin and evolution of this
binary system (see Possenti 2002; Orosz \& van Kerkwijk 2003; Grindlay et al.
2002 for a review).

In the framework of a coordinated spectro-photometric project devoted
to investigate the nature of this peculiar object, we recently
reported on preliminary results from Very Large Telescope UVES high
resolution observations: we measured the mass ratio of the system
($q=M_{\rm PSR}/M_{\rm COM}=5.85\pm 0.13$) with unprecedented
precision for a non-relativistic binary pulsar system and confirmed
the cool effective temperature ($\sim$5,500 K) of \com~(Ferraro et
al. 2003). A radial velocity $V_{\rm bary}=17.7\pm 2.3$ km s$^{-1}$
was also derived for the binary barycenter, fully compatible with the
systemic radial motion of the cluster (Harris 1996).

Here we focus on a detailed analysis of the H$\alpha$ and He~{\sc i}
lines: the complex structure of the former is used to investigate the
distribution of matter lost by \com, whereas the study of the He~{\sc
i} feature aims at confirming the hypothesis that a narrow, hot strip
(T$_{eff}>$10,000~K) is located on the pulsar companion surface
(Ferraro et al.  2003), and at estimating the area of the heated
region.

\section{The H$\alpha$ region}
\label{Halfa}

At radio wavelengths, \psr~displays eclipses for more than 40\% of its orbital
period, and evident irregularities of the signal at all orbital phases. The
latter characteristic was ascribed by D'Amico et al. (2001b) to the motion of
the MSP within a large envelope of matter released by the secondary. This
hypothesis finds additional support in the modulated X-ray emission detected by
{\it Chandra} (Grindlay et al. 2001, 2002), which may be due to the interaction
between the MSP wind and the stellar matter. Furthermore, the anomalous
position of \com~in the (V, H$\alpha-$R) color-magnitude diagram (see Fig.~1 of
Ferraro et al. 2001  and Taylor et al. 2001) suggests a  H$\alpha$ excess:
this is a unique feature for a MSP companion  and could  suggests either 
RS~CVn--like activity in the companion atmosphere (Orosz \& Van Kerwijk 2003)
or the presence of significant interactions between the energetic pulsar flux
and the matter lost by \com.

Motivated by these observations, we have exploited high quality, phase
resolved spectroscopic data to carefully look at the H$\alpha$ line
region. A set of 8 high resolution spectra of \com~was obtained with
the {\it Ultraviolet-Visual Echelle Spectrograph} (UVES) mounted on
the ESO Very Large Telescope on Cerro Paranal (Chile) in 2002
June--July.  The observational strategy and the data are described in
Ferraro et al. (2003; see their Table~1).  The system hosting \psr~is
located in a crowded region of the cluster; thus, in order to avoid
possible contamination from nearby stars, we have here used only
spectra acquired in the best seeing conditions, which correspond to
orbital phases $\phi \sim$ 0.02, 0.56, 0.78, and 0.36, close to
quadratures, opposition and conjunction, respectively ($\phi=0.25$
when \com~is at the minimum distance from the observer).

As can be seen in Figure 1 (right panel), \com~spectra clearly show
H$\alpha$ absorption lines, but the line profiles appear complex
and require a detailed analysis. To perform this, the
\com~spectra must be compared with a reference template. We took
advantage of the availability of spectra observed with a similar UVES
configuration in a project devoted to study the chemical abundance in
NGC~6397 (Gratton et al. 2001). In particular the high quality spectra
of three subgiants (namely \# 669, 793 and 206810 in Gratton et al.)
were averaged and convolved with a rotation profile of $V_{\rm
rot}=49.6$ km s$^{-1}$ in order to account for the inferred rotation
velocity of \com~(Ferraro et al. 2003).  The resulting spectrum was
then normalized to the \com~continuum at each orbital phase and used
as a template. The temperatures of the three subgiants are fully
comparable with that derived for \com~($\sim$ 5,500 K, Ferraro et
al. 2003; Orosz \& van Kerkwijk 2003) and their overall chemical
composition is found compatible with that observed in \com~(although
with a few notable exceptions, as will be discussed in Sabbi et
al. 2003, in preparation). In fact, the agreement between the template
and the \com~spectra in the H$\gamma$ region is excellent (see the
left panel of Fig.~1), supporting the similarity of \com~with
respect to a normal subgiant branch star of NGC~6397 and hinting at
the power of the differential analysis.

By using this method, we can easily pinpoint the anomalies of the
\com~spectra in the H$\alpha$ region (right panel of Fig.~1): {\it
(i)} the line shallowness, which suggests a certain amount of
H$\alpha$ emission, and {\it (ii)} the complex structures in the line
wings. In order to better analyze these features, we subtracted the
template spectrum from the \com~one. The result for the spectrum taken
at $\phi=0.56$ is presented in Figure~2, whereas Figure~3 displays the
changing structure of the H$\alpha$ emission line along the orbit. At
least two different H$\alpha$ emission components can be
distinguished. The first one appears as a bright, narrow and symmetric
line, clearly visible as the main peak at $\lambda_p\sim 6567$ {\AA}
in Figure~2. At all orbital phases the narrow peak is always easily
recognizable (see Fig.~3). The second component is much fainter,
broader and asymmetric, characterized by a large bump at
$\lambda<\lambda_p$ and a long tail at $\lambda>\lambda_p$ in
Figure~2. It covers a large velocity range: in the highest quality
available spectra (taken at phase $\phi=0.56$), its peak displays a
slow radial velocity of $V_{\rm b}\sim 18$ km$^{-1}$ (in the observer
frame, see Fig.~2), while the terminal velocity of the tail $V_{\rm
t}$ is about 20 times higher. The overall shape of this broad second
component appears (specularly reflected with respect to the shape
observed at $\phi=0.56$) also at the opposite quadrature (panels $d$
and $d\arcmin$ of Fig.~3), whereas near opposition and conjunction
only the weakest tail is distinguishable (panels $a$, $a\arcmin$, $c$
and $c\arcmin$ of Fig.~3).

Since the differential analysis clearly reveals H$\alpha$ emission
while excluding any significant H$\gamma$ emission (Fig.~1, left
panel), we can conclude that there is a large decrement in Balmer line
intensity and hence that the gas emitting the H$\alpha$ photons must
be optically thin.

\section{Helium lines}  
\label{He}

As first reported by Ferraro et al. (2003), strong He~{\sc i} absorption lines
are present in most of the observed spectra, implying that a region of the
\com~surface has a high temperature, T$_{eff}>$10,000 K. This feature was
unexpected since the other temperature indicators (e.g., photometric colors and
H$\alpha$ wings) suggest that the surface temperature is $\sim$ 5,500 K
(Ferraro et al. 2001; Orosz \& van Kerkwijk 2003). On the basis of a
qualitative inspection of the He~{\sc i} line widths, Ferraro et al. (2003)
proposed the existence of a hot strip along the \com~surface.

In the aim of better investigating this hypothesis, we carefully compared
the shape and the broadening profile of the He~{\sc i} lines with
those of other \com~spectral lines. In particular we have built 4
empirical rotation line profiles based on many lines in the blue part
of the spectrum, and convolved these profiles with that of the He~{\sc
i} lines. The comparison with the observed spectrum (Fig.~4) shows
that the empirical profiles can nicely reproduce both the Doppler
shift and the rotational profiles of the observed He~{\sc i} lines at
all the orbital phases. As a consequence, the hot region must be
longitudinally extended across the stellar surface.  Moreover, at
phase $\phi$=0.36 (i.e. when the surface of the companion facing the
pulsar is almost invisible to the observer) the He~{\sc i} line
appears much fainter than at the other orbital phases. 
%and its width is smaller than the empirical rotation profile. 
We can conclude that the hot region does not encircle the entire star, but is
primarily located on the companion hemisphere visible from the pulsar.

One can estimate the projected area of such a heated region. For this purpose,
we assumed a simple model where the atmosphere of the star is composed of two
(horizontal) regions: one hot, with temperature roughly similar to that of a B
star; and one cool, with the temperature of a G8 star.  The ratio between the
fluxes emitted by these two regions can be estimated by comparing the residual
emission at the center of the Ca~{\sc i}~K line at 3933 \AA~(EW = 4.1\AA)
where only light from the cool region is seen and at the center of the He~{\sc
i} line at 5878 \AA~(EW = 0.19\AA) where we assume to be observing only
light from the hot region. The flux ratios emitted at these two wavelengths by
stars of spectral type B7 (T$_{\rm eff}=15,000$ K) and G8 (T$_{\rm eff}=5,500$
K) were taken from Gunn \& Stryker (1983). Comparing these ratios with the
\com~spectrum, suggests that the heated region should cover less than 1\% of
the stellar surface.

\section{Discussion}  

\subsection{H$\alpha$ lines}
\label{disH}

Figure~3 (left panel) shows the subtracted spectra corrected for the Doppler
shift due to the orbital motion of \com~around the MSP: since at each orbital
phase the main peak has always the same radial velocity as \com, we can safely
conclude that the brightest peak is due to gas emitted by a hot and optically
thin (i.e. rarefied) gaseous region close to the star surface, 
%namely the stellar chromosphere.
as often observed in the atmospherically active binaries (e.g. Montes et
al. 1997,2000 for a review).

The evident asymmetric shape (Fig.~2) of the secondary broad component
of the H$\alpha$ emission precludes an origin in a Keplerian disk
surrounding \psr. Instead, the synchronous rotation --- modulated by
the radial orbital velocity around the center of mass of the system
--- of the broad and the narrow components (particularly evident when
comparing panels {\it b'} and {\it d'} of Fig.~3) is suggestive of a
lump of matter protruding from \com~and roughly orbiting as a rigid
body with the companion star. In this picture, the broad peak of the
fainter component (corresponding in quadrature to a radial velocity
$(V_{\rm b}-V_{\rm bary})\sim 0$ with respect to the center of mass of
the binary) would represent hydrogen gas flowing from the Roche lobe
of \com~toward the center of mass of the system (located at $\sim
1~{\rm R_\odot}$ from the pulsar), where it is stopped and then swept
away by the pulsar energetic flux of radiation and/or accelerated
particles (a similar {\it radio-ejection}
mechanism has been postulated for this binary system by Burderi, D'Antona \&
Burgay 2002). The swept material could form a wide (cometary-like?)
tail possibly winding around the binary. The portion of the gaseous
tail which is closest to \com~should follow almost rigidly the orbital
motion of the companion star, always pointing roughly in the opposite
direction (with respect to \com) of the MSP. If it produces the
declining end of the broad line component (see Fig. 2), the terminal
radial velocity seen at quadratures ($V_{\rm t}\sim 300-400$ km
s$^{-1}$) would imply that the swept gas maintains a high enough
temperature for emitting H$\alpha$ photons up to $R_{\rm t}=(V_{\rm
t}P_{\rm orb}/2\pi-a_{\perp}q)/\sin i \sim (5.2~{\rm R_\odot})/\sin i$
outside the orbit of \com, where $P_{\rm orb}$ and $a_{\perp}$ are the
orbital period and the projected semimajor axis of the pulsar orbit
(D'Amico et al. 2001b), whereas $q$ and $i (\sim 50~\deg$; Ferraro et
al. 2003) are the mass ratio and the orbital inclination.

Interestingly, $R_{\rm t}$ is of the same order of the radius of the
region (concentric with the companion star) causing the long eclipses
of \psr~at 1.4 GHz (D'Amico et al. 2001b). Thus the gas from which the
H$\alpha$ broad emission originates may be responsible also for the
radio eclipses of \psr~(at 1.4 GHz) between orbital phases 0.05
and 0.45. The matter blown away from the binary at larger distances
than $R_{\rm t}$ is probably cooler and not completely ionized. It
would be undetectable at optical wavelengths, but could cause the
striking irregularities displayed by the radio signal of \psr~
%even at orbital phases far from the nominal eclipse region 
at nearly all orbital phases, even far from the nominal eclipse region
(D'Amico et al. 2001b).

\subsection{He~{\sc i} lines}

In order to understand the origin of the He~{\sc i} lines, we looked
for other systems that show physical and/or spectral
similarities. AM~Her binaries are a possibility, since they contain a
WD and a MS star, and do not have an accreting disc, since the
magnetic field stops its formation; but the spectra of these stars are
dominated by high excitation emission lines such as He~{\sc ii} (e.g.,
Schwope et al. 1997), while we do not find these features in our
spectra. The Herbig~Ae/Be stars, young intermediate mass stars, have
similar spectra which show H$\alpha$ emission and He {\sc i}
absorption lines (see Waters \& Waelkens 1998 for a review). But even
if the origin of the H$\alpha$ lines is similar, the T$_{\rm eff}$ of
\com~ is too low to justify the helium. We have to consider other
mechanisms to justify the helium presence in our spectra.

The detailed inspection of the rotation profile and of the orbital
modulation of the He~{\sc i} absorption lines strongly supports the
existence of a hot ``barbecue-like'' strip{\footnote{A suggestive
definition proposed by Josh Grindlay (private communication).}}
elongated across the stellar photosphere facing the pulsar. Assuming
for simplicity that a rectangular ribbon encircles $\sim$75\% of the
star (in order to accomplish the observed weakening of the He~{\sc i}
absorption line near only one of the four orbital epochs) and has an
area $\lapp 1$ \% of the star photosphere (as estimated in
\S\ref{He}), the ribbon's length would be $\sim 7~{\rm R_\odot}$ and
its width $\lapp 0.04~{\rm R_\odot}.$

What can cause such a ``stretched feature''? The high temperatures
required for producing the He~{\sc i} absorption line could be due to
a shocked filament standing, e.g., at the boundary between the
companion surface and the gas swept back toward the companion by the
pulsar flux. However it seems difficult to maintain a hot filament of
gas confined within a few hundredths of ${\rm R_\odot}.$ Another
intriguing possibility is that the heated strip is due to the effects
of the pulsar irradiation on the companion surface. In this case, the
shape of the heated region can be accounted for only by a highly
anisotropic pulsar emission pattern confined (at least in the
direction orthogonal to the binary orbital plane) within a
surprisingly small angle of $\sim 0.4$ degrees.  Since such a narrow
pencil (or highly flattened fan) beam would probably illuminate the
\com~ surface only for a portion of the orbit, the tiny width of the
heated region would be preserved if the cooling time of the irradiated
gas is much shorter than the orbital timescale $\sim 1$ day. Under
basic assumptions (as supposing a perfect gas and an atmospheric
density of $10^{-7}$ g cm$^{-3}$), we estimate that only a few hundreds
seconds are necessary to cool the gas from 10,000 K to 5,500 K.

Finally we note that the occurrence of highly anisotropic pulsar
emission would explain why irradiation does not leave any detectable
imprint on the photometric curves of \com~presented by Orosz \& van
Kerkwijk (2003) and by Kaluzny et al. (2003), without the need for
invoking any reduction of the intrinsic spin period derivative $\dot P$ of
\psr~with respect to the value calculated using radio timing data
(D'Amico et al. 2001b).

\acknowledgements{\small We thank the referee Peter Edmonds
 for his helpful 
suggestions, which improved the presentation.
Financial support to this research has been
provided by the Agenzia Spaziale Italiana (ASI) and the {\it Ministero
dell'Istruzione, dell'Universit\`a e della Ricerca} (MIUR).  FC is
supported by the US NSF and NASA.}

\clearpage

\begin{figure} 
\plotone{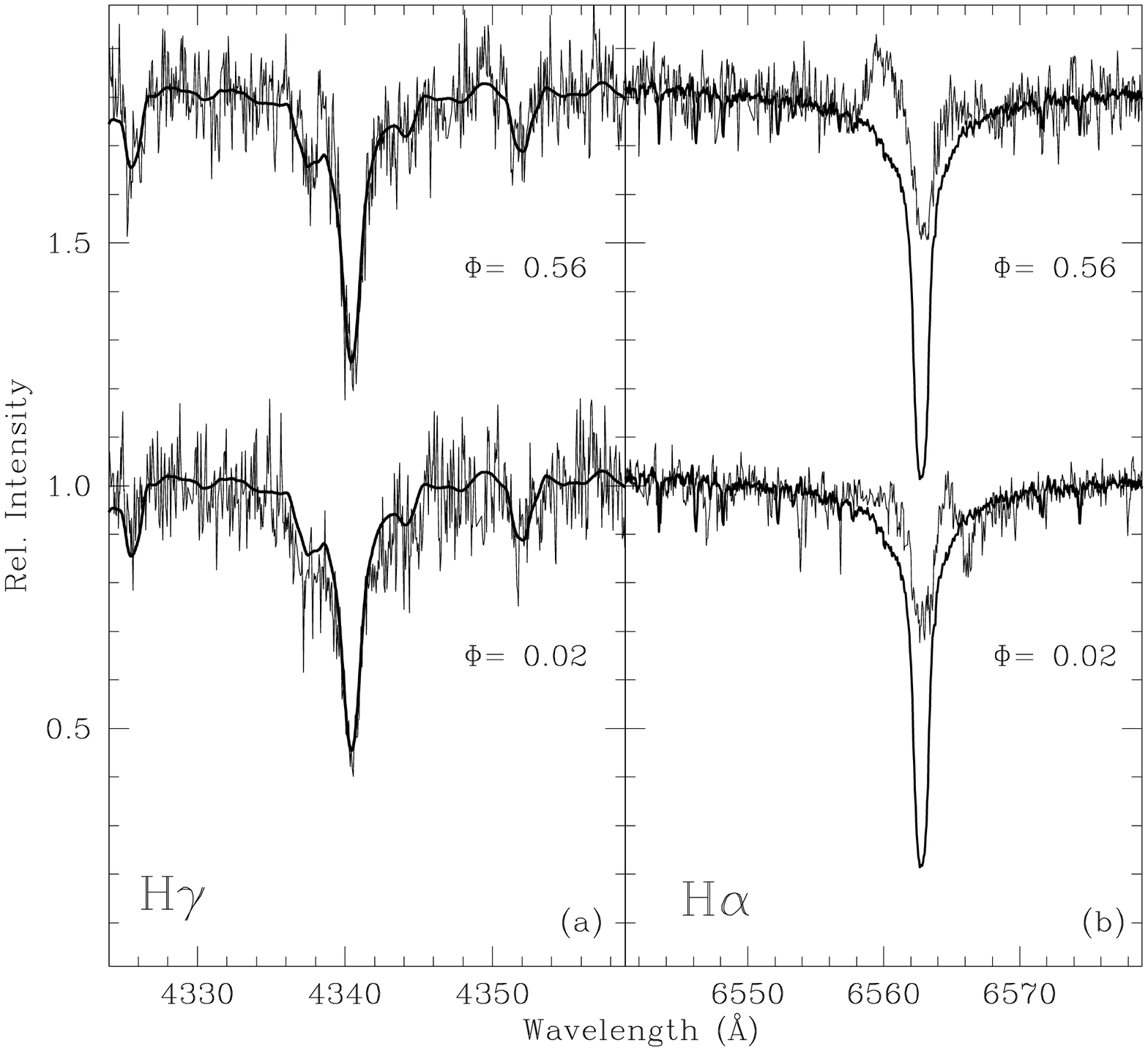}
\caption {
The observed spectra ({\it light solid lines}) of \com~in the H$\gamma$ ({\it
left panel}) and H$\alpha$ ({\it right panel}) regions at two orbital phases
($\phi$=0.02 and $\phi$=0.56, respectively.  We have applied the
same convention of D'Amico et al. 2001b, in which \com~is nearest to the
observer at $\phi$=0.25). Template spectra (heavy solid lines), obtained by
averaging the spectra of three normal subgiants of NGC~6397 (Gratton et al.
2001), are overplotted for comparison: note the excellent agreement between the
spectra near H$\gamma$; conversely the H$\alpha$ absorption in \com~is clearly
shallower and shows anomalous wings, suggesting the existence of a complex
emitting structure.} \end{figure}

\clearpage

\begin{figure} 
\plotone{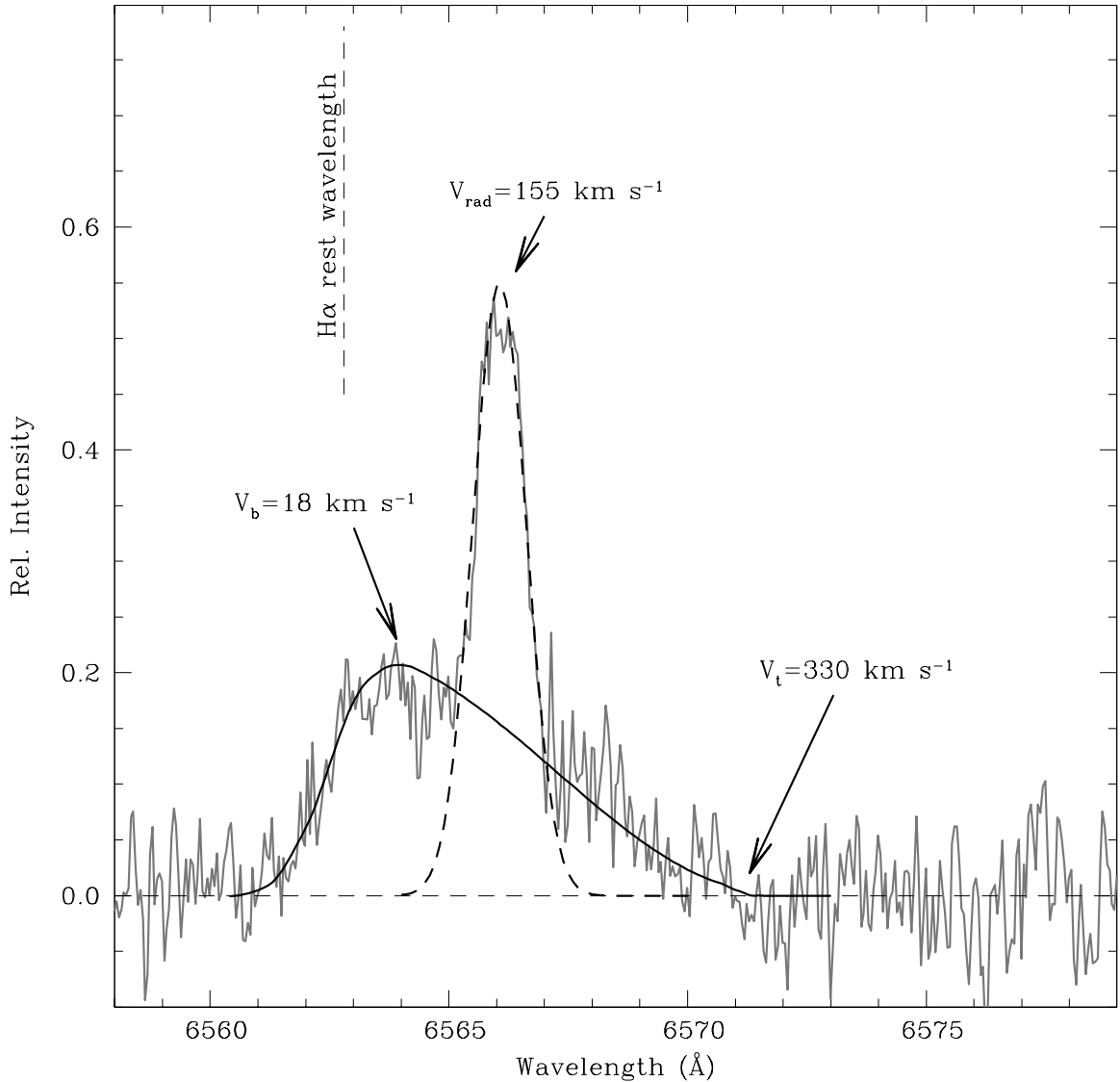}
\caption{\label{Hemisdet} 
Subtracted spectrum of \com~(see  text for description of the adopted
procedure) in the H$\alpha$ region at orbital phase $\phi=0.56$.  No
wavelength shift has been applied to the spectrum. The {\it vertical dashed
line} marks the H$\alpha$ line rest wavelength and the {\it horizontal dashed
line} is the continuum level.  Two components can be seen: the narrower and
brighter (modeled by the {\it heavy dashed line}) is emitted by the stellar
chromosphere (see \S\ref{disH}) while the broader and fainter (modeled by the
{\it heavy solid line}) shows a peak (here named $V_b$) at smaller velocities
than the star, but extending farther, with the velocity of its tail ($V_t$)
greater than 300 km s$^{-1}$.}
\end{figure}

\clearpage

\begin{figure} 
\plotone{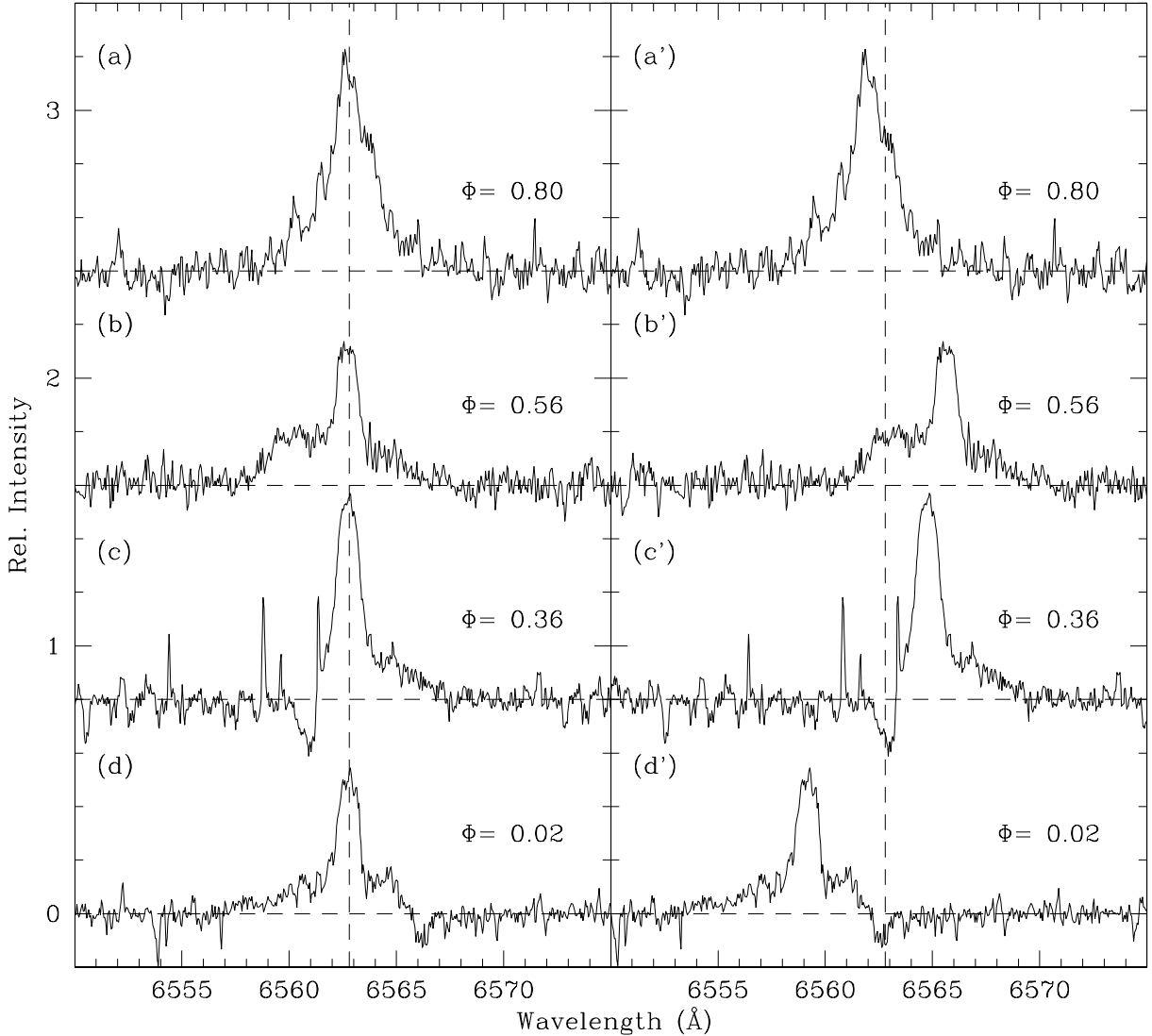}
\caption{\label{Hemis1} 
Subtracted spectra of \com~at 4 orbital phases. In the {\it left panel} the
Doppler shifts corresponding to the orbital motion of \com~(Ferraro et al. 
2003) are applied, while  in the {\it right panel} a Doppler shift 
corresponding to the center of gravity radial velocity has been applied to 
each spectrum. The spectra are arbitrarily shifted in intensity for the sake
of clarity. The {\it horizontal dashed lines} are the continuum levels, while
{\it the vertical dashed lines} represent the H$\alpha$ line rest wavelength.
Note that in the left panel the main peak, at each phase, has the same radial
velocity as \com, denoting its chromospheric origin.}
\end{figure}

\clearpage

\begin{figure} 
\plotone{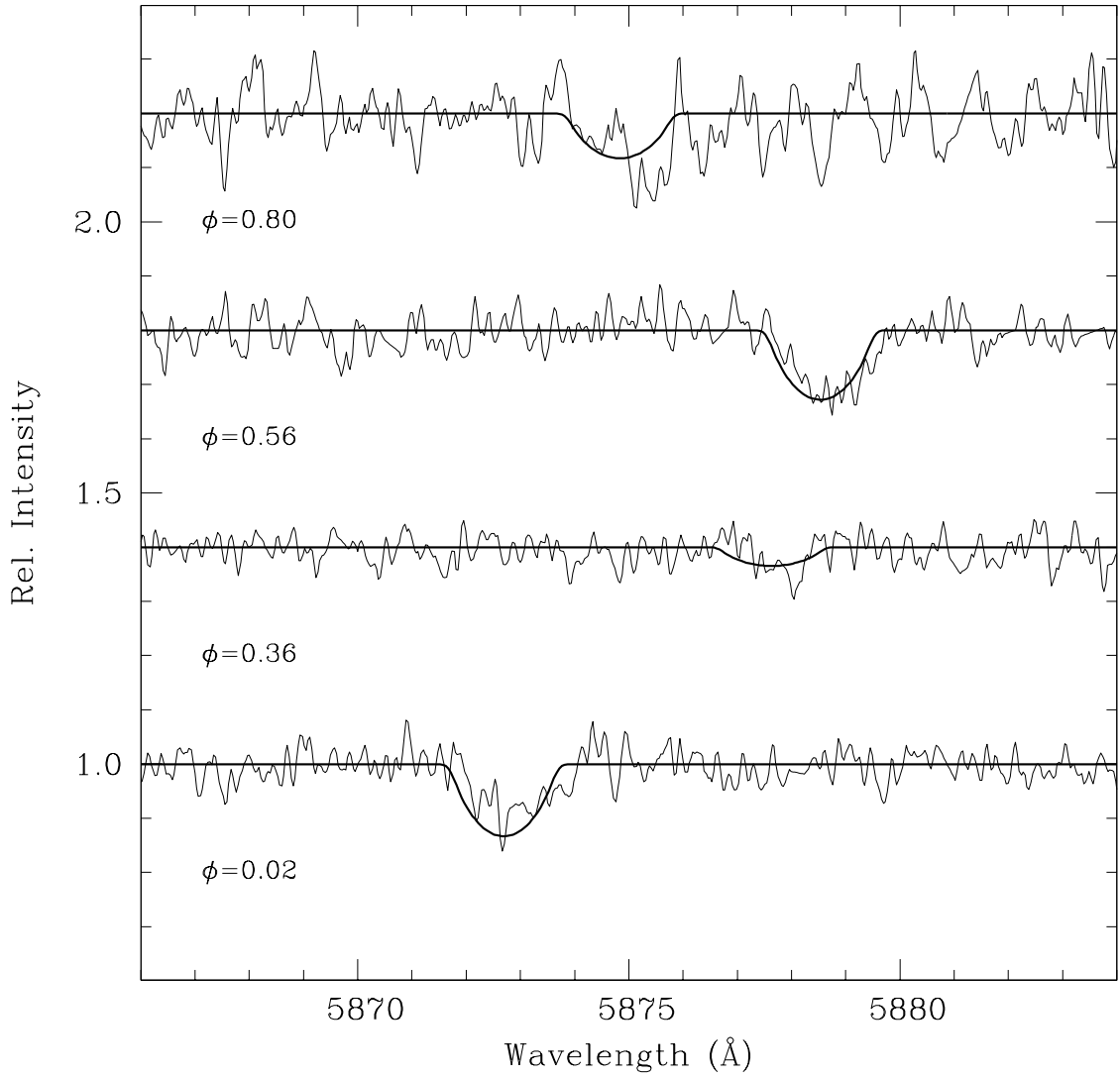}
\caption{\label{He-line} 
The heavy solid lines overplotted on the observed spectra are
empirical profiles obtained by combining the typical rotation
broadening profiles with the He {\sc i} line intensity.  An arbitrary
shift in intensity is applied to the spectra for the sake of clarity.}
\end{figure}

\clearpage


\begin{thebibliography}{}

\bibitem[Burderi, D'Antona \& Burgay 2002]{bdb02}
Burderi, L., D'Antona. F. \& Burgay, M. 2002, ApJ, 574, 325

\bibitem[D'Amico et al. 2001a]{dlm+01}
D'Amico, N., Lyne, A. G., Manchester, R. N., Possenti, A. \& Camilo, F. 2001a,
ApJ, 548, L171

\bibitem[D'Amico et al. 2001b]{dpm+01}
D'Amico, N., Possenti, A., Manchester, R. N., Sarkissian, J., Lyne, A. G.\&
Camilo, F. 2001b, ApJ, 561, L89

\bibitem[Ferraro et al. 2001]{fpds01}
Ferraro, F.~R., Possenti, A., D'Amico, N. \& Sabbi, E. 2001, ApJ, 561, L93 

\bibitem[Ferraro et al. 2003]{fpds03}
Ferraro, F.~R., Sabbi, E., Gratton, R., Possenti, A., D'Amico, N., 
Bragaglia, \& A., Camilo, F. 2003, ApJ, 584, L13

\bibitem[Gratton et al. 2001]{g+01}
Gratton, R. et al. 2001, A\&A, 369, 87

\bibitem[Grindlay et al. 2001]{ghemc01}
Grindlay, J.~E., Heinke, C.~O., Edmonds, P.~D., Murray, S., \& Cool, A. 2001, 
ApJ, 563, L53

\bibitem[Grindlay et al. 2002]{grindlay02} 
Grindlay, J.~E., Camilo, F., Heinke, C.~O., Edmonds, P.~D., 
Cohn, H., \& Lugger, P. 2002, ApJ, 581, 470

\bibitem[]{}
Gunn, J.~E. \& Stryker, L.~L. 1983, ApJS, 52, 121

\bibitem[Harris 1996]{h96}
Harris, W~E. 1996, AJ, 112, 1487

\bibitem[Kaluzny et al. 2003]{kaluzny03} 
Kaluzny, J., Rucinski, S.~M., Thompson, I.~B. 2003, AJ, 125, 1546

\bibitem[Montes et al. 2000]{M00}
 Montes, D., Fern\'andez--Figueroa, M.~S., De Castro, E., Cornide, M.,
Latorre, A., \& Sanz--Forcada, J. 2000, A\&AS, 146, 103


\bibitem[Montes et al. 1997]{M97}
 Montes, D., Fern\'andez--Figueroa, M.~S., De Castro, E., \& Sanz--Forcada,
J. 1997, A\&AS, 125, 263

\bibitem[Orosz \& van Kerkwijk 2003]{ov03}
Orosz, J.~A., \& van Kerkwijk, M.~H. 2003, A\&A, 397, 237

\bibitem[Possenti 2002]{p02}
Possenti, A. 2002, Neutron Stars Pulsars and Supernova Remnants,
eds. W. Becker, H. Lesch \& J. Tr\"umper, MPE Report 278, p. 183

\bibitem[Schowpe, Mantel \& Horne 1997]{smh97}
Schwope, A.~D., Mantel, K.~H., \& Horne, K. 1997, A\&, 319, 894

\bibitem[Taylor et al. 2001]{t2001}
Taylor, J.~M., Grindlay, J.~E., Edmonds, P.~D., Cool, A.~M. 2001, ApJ, 553,
L169

\bibitem[Waters \& Waelkens 1998]{ww98}
Waters, L.~B.~F. \& Waelkens, C. 1998, ARAA, 36, 233

\end{thebibliography}
\end{document}